# Constructing the body source for high-order lattice Boltzmann method

Shunyang Li [1], Li Wan [1], Nan Gui [1], Xingtuan Yang [1], Jiyuan Tu [1,2],

Shengyao Jiang [1]

1. Institute of Nuclear and New Energy Technology, Collaborative Innovation Center of Advanced Nuclear Energy Technology, Key Laboratory of Advanced Reactor Engineering and Safety, Ministry of Education, Tsinghua University, Beijing, 100084, China

2. School of Engineering, RMIT University, Melbourne, VIC 3083, Australia

**Abstract** This paper presents a novel strategy for constructing body source terms in the high-order lattice Boltzmann method (LBM), designed to efficiently introduce various physical phenomena by modifying the non-equilibrium distribution function. The source term, expressed as a Hermite polynomial, provides a flexible framework for simulating complex fluid flows. Three typical source terms are given: a body force source for gravity-driven flows, a thermal dissipation source for controlling the Prandtl number, and a pressure tensor source for modeling multiphase flows. Chapman-Enskog analysis confirms that the source terms recover the expected macroscopic equations. Notably, the proposed strategy eliminates the need for explicit construction of the collision operator, a challenge in conventional approaches for handling diverse physical scenarios. Furthermore, the method is compatible with the traditional BGK model, ensuring its applicability to various high-order lattices. The model's accuracy and versatility are validated through a series of benchmark tests, showing excellent agreement with existing literature results.

# 1. Introduction

The lattice Boltzmann method (LBM) has been attracting considerable interest for its high efficiency and ease of parallelism. It has been successfully applied in a broad area [1] [2] [3] [4] [5]. The standard LBM solves the Boltzmann equation with a Bhatnagar-Gross-Krook (BGK) collision operator [6]. The hydrodynamic parameters are inferred from the particle distribution function after the collision-streaming step. According to Grad's theory [7], a fluid state is fully characterized by the zeroth through third-order velocity moments of the distribution function. However, the standard LBM are limited to the first three order moments, primarily due to its athermal nature where temperature is excluded. In thermal applications, temperature is often computed separately using another equation, such as in the double distribution function method [8] [9]. Nevertheless, neglecting high-order moments in the standard method leads to certain limitations. The Chapman-Enskog analysis reveals that truncating the equilibrium distribution function to the second order generates an error term, $\nabla \cdot (\rho \boldsymbol{uuu})$, in the momentum equation, thereby violating the Galilean invariance at high Mach numbers. Linear stability analysis indicates that the standard LBM becomes unstable when Ma exceeds 0.731 [10]. Various approaches have been proposed to mitigate this deficiency by refining the relaxation of velocity moments [11-13]. Nevertheless, such methods often fail to exert full control over high-order moments. Additionally, standard LBM suffers from insufficient lattice isotropy [14], which results in non-physical artifacts such as spurious currents in multiphase flow simulations [15]. The isotropy deficiency requires correction terms, further complicating the model [16] [17]. Thus, although the high-order moments do not explicitly appear in traditional LBM formulations, their influence on simulation accuracy is non-negligible.

The high-order LBM is initially introduced for simulating thermal compressible flows [18] [19], as it retrieves the Navier-Stokes-Fourier equation. This approach solves both hydrodynamic and thermodynamic variables within a single lattice Boltzmann equation. The equilibrium distribution function is extended with high-order polynomial terms [18,20]. Philippi et al. [14] proposed a multi-speed lattice for LBM with enhanced quadratic accuracy for high-order moment calculation and improving isotropy. The space-filling discrete velocities make the multi-speed lattice method more competitive against off-lattice methods. Shan et al. [21] simplified the derivation of high-order LBM by projecting the distribution function onto a space spanned by Hermite polynomials. Besides multispeed LBM, other high-order schemes have been developed. Frapolli et al. [22] presented an entropic multispeed lattice Boltzmann method where moments are relaxed by maximizing the entropy. Atif et al. proposed a body-centered-cubic (bcc) lattice that improves spatial accuracy through the inclusion of extra lattice points [23]. Geier et al. developed the central moment method with moments of different orders evolved separately [24]. Coreixas et al. proposed a regularization step aimed at filtering undesirable high-order moments during the collision step, therefore minimizing their impact on the solution [25].

The high-order LBM retains most of the benefits of the standard LBM while offering unique advantages. Firstly, it eliminates the momentum error $\nabla \cdot (\rho \boldsymbol{uuu})$, thereby preserving Galilean invariance [26]. Secondly, the high-order LBM exhibits better stability by explicitly incorporating control over high-order velocity moments. Siebert et al. [27] demonstrated that the high-order equilibrium distribution function provides greater stability than conventional second-order formulation. Renard et al. reported that the maximal allowable Mach number for high-order LBM can extend up to 1.1 [28]. Additionally, the integration of high-order moments makes the high-order LBM suitable for simulating flows with complex intermolecular interactions. When simulating thermal compressible flows, the mass, momentum, and energy are inherently coupled in one equation. Therefore, additional coupling techniques are no longer needed. Recent studies suggest the high-order LBM is also promising for microflows [29], rarefied gas flows [30] and supersonic flows [22].

Despite its benefits, the application of high-order LBM is constrained by certain challenges. The increased complexity of the collision operator is particularly evident when incorporating high-order moments. For example, in the multispeed LBM, the Prandtl number is fixed to unity if all moments are relaxed at an identical rate. The issue is commonly addressed by relaxing moments at different rates or by reconstructing the collision operator entirely [31] [32]. Although these strategies recover heat dissipation with arbitrary thermal conductivity. However, these approaches often complicate the model's implementation. For more intricate interactions such as multiphase flows, constructing an appropriate collision operator becomes even more challenging.

An alternative strategy to address these challenges involves the use of source terms to introduce additional effects within the LBM framework. In this approach, the BGK operator focuses on recovering the viscosity stress, while other effects are incorporated through source terms. The source term directly modifies the non-equilibrium component of the distribution function, avoiding the need for complex kinetic collision operators. The method has proven effective in standard LBM. Guo et al. used source terms to incorporate body forces into the momentum equation [33]. Malaphis et al. used source terms to construct a subgrid tensor for LES simulation [34]. Renard et al. used source terms to adjust the viscous stress tensor for arbitrary specific heat capacities [35]. However, source terms are typically tailored for specific applications, limiting their applicability across different scenarios. For high-order LBM, it is crucial to consider the impact of the source term on high-order velocity moments. Thus, developing a generalized procedure for constructing source terms is essential to extend the applicability of high-order LBM.

The study aims to develop a systematic approach to generate source terms in high-order LBM. To clarify this problem the paper is organized as follows. Section 2 introduces the fundamentals of the high-order lattice Boltzmann method and illustrates the derivation of source terms based on Chaplan-Enskog analysis. Section 3 presents several representative source terms developed through this framework, followed by the specific implementation for multispeed lattice in Section 4. Section 5 presents numerical validations on thermal compressible flows and multiphase flows. Section 6 is the conclusion.

## 2. The high-order LBM with source term

## 2.1 The lattice Boltzmann equation

The lattice Boltzmann equation describes the behavior of the particle distribution function $f$. Its continuum form is expressed as [21]

$$\frac{\partial}{\partial t}f_i + \pmb{\xi}_i \cdot \frac{\partial}{\partial \pmb{x}}f_i = \Omega_{\text{BGK}}(f_i) + S_i, \qquad (2.1)$$

where $\pmb{\xi}_i$ is the quadrature point in the continuous velocity space, and $\Omega_{\text{BGK}}(f_i) = -1/\tau(f_i - f_i^{eq})$ is the BGK collision operator, with relaxation time $\tau$ determining the rate at which the system evolves to its equilibrium. The equilibrium distribution function $f_i^{eq}$ can be written as [34]

$$\begin{aligned} f_i^{eq,4th} = w_i \rho \Big\{ & 1 + \pmb{\xi}_i \cdot \pmb{u} + \frac{1}{2}(\pmb{\xi}_i \cdot \pmb{u})^2 - \frac{1}{2}\|\pmb{u}\|^2 + \frac{1}{2}\bar{\theta}\big(\|\pmb{\xi}_i\|^2 - D\big) \\ & + \frac{1}{6}(\pmb{\xi}_i \cdot \pmb{u})^3 - \frac{1}{2}(\pmb{\xi}_i \cdot \pmb{u})\|\pmb{u}\|^2 + \frac{1}{2}(\pmb{\xi}_i \cdot \pmb{u})\bar{\theta}\big[\|\pmb{\xi}_i\|^2 - (D+2)\big] \\ & + \frac{1}{24}(\pmb{\xi}_i \cdot \pmb{u})^4 - \frac{1}{4}(\pmb{\xi}_i \cdot \pmb{u})^2\|\pmb{u}\|^2 + \frac{1}{8}\|\pmb{u}\|^4 \\ & + \frac{1}{4}\bar{\theta}\left[(\pmb{\xi}_i \cdot \pmb{u})^2\big(\|\pmb{\xi}_i\|^2 - (D+4)\big) + \|\pmb{u}\|^2\big((D+2) - \|\pmb{\xi}_i\|^2\big)\right] \\ & + \frac{1}{8}\bar{\theta}^2\big[\|\pmb{\xi}_i\|^4 - 2(D+2)\|\pmb{\xi}_i\|^2 + D(D+2)\big] \Big\} \end{aligned}, \qquad (2.2)$$

where $\rho$ is the density, $\pmb{u}$ is the velocity. $\bar{\theta} = \theta - 1$, where $\theta$ is the temperature. The high-order terms account for thermal and compressible effects in the fluid. The equilibrium distribution function reflects a steady state of the fluid system subject to given macroscopic parameters regardless of position or time.

All variables shown in the equations are dimensionless, scaled by appropriate reference values for length, density, and temperature scale $L_r$, $\rho_r$ and $T_r$. The velocity scale $u_r$ is not independent in thermal compressible flows, as it is given by $u_r = \sqrt{R_g T_r}$, where

$R_g$ is the gas constant.

The macroscopic quantities of the fluid, such as mass, momentum, temperature, and stress are derived from the velocity moments of the distribution function. The $k$-th order velocity moment $\mathcal{F}_k$ is defined as $\mathcal{F}_k = \sum_i f_i \boldsymbol{\xi}^k$. The state of fluid can be fully described if $\mathcal{F}_0$, $\mathcal{F}_1$, $\mathcal{F}_2$ and $\mathcal{F}_3$ are known [7]. These moments must satisfy the corresponding relations for conservation, i.e.,

$$\mathcal{F}_0 = \mathcal{F}_0^{eq}, \mathcal{F}_1 = \mathcal{F}_1^{eq}, \text{tr}(\mathcal{F}_2) = \text{tr}(\mathcal{F}_2^{eq}), \tag{2.3}$$

where tr($\cdot$) is the trace of the tensor.

## 2.2 The continuum limit

The source term $S_i$ in Eq. (2.1) accounts for external forces or other macroscopic effects that influence the distribution function. In many cases, the mesoscopic behavior of the distribution function is not explicitly known, but its impact on the macroscopic variables is understood. This section aims to express the source term $S_i$ in terms of macroscopic quantities.

Upon projecting the lattice Boltzmann equation onto the Hermite polynomial basis, the resulting equations correspond to mass, momentum, and stress conservation laws. Therefore, the source term $S_i$ can be written as a series expansion in Hermite polynomials:

$$S_i = w_i \sum_{n=0}^{\infty} \frac{1}{n!} \boldsymbol{s}^{(n)} \cdot \boldsymbol{\mathcal{H}}^{(n)}(\boldsymbol{\xi}_i), \tag{2.4}$$

where $\boldsymbol{\mathcal{H}}^{(n)}(\boldsymbol{\xi}_i)$ denotes the $n$th-order Hermite polynomial $\boldsymbol{\mathcal{H}}^{(n)}(\boldsymbol{\xi}_i) = \boldsymbol{\xi}_i^n - [\boldsymbol{\delta}\boldsymbol{\xi}_i^{n-2}] + [\boldsymbol{\delta}^2 \boldsymbol{\xi}_i^{n-4}] - \cdots$, and $\boldsymbol{s}^{(n)}$ are the corresponding macroscopic coefficients.

For clarity the tensor product notation between the isotropy tensor $\boldsymbol{\delta}$ and $\boldsymbol{\xi}_i$ is omitted. The bracket [] denotes a summation constructed by a permutation of indices, i.e., $[\boldsymbol{\delta}\boldsymbol{\xi}]_{\alpha\beta\gamma} = \delta_{\alpha\beta}\xi_\gamma + \delta_{\alpha\gamma}\xi_\beta + \delta_{\beta\gamma}\xi_\alpha$.

The discrete distribution function $f_i$ can be written as

$$\begin{aligned} f_i &= w_i \sum_{n=0}^{\infty} \frac{1}{n!} \boldsymbol{a}^{(n)} \cdot \boldsymbol{\mathcal{H}}^{(n)}(\boldsymbol{\xi}_i) \\ &= w_i \sum_{n=0}^{\infty} \frac{1}{n!} \left(\boldsymbol{a}^{(n,eq)} + \boldsymbol{a}^{(n,neq)}\right) \cdot \boldsymbol{\mathcal{H}}^{(n)}(\boldsymbol{\xi}_i), \end{aligned} \qquad (2.5)$$

where $\boldsymbol{a}^{(n,eq)}$ and $\boldsymbol{a}^{(n,neq)}$ are the equilibrium and non-equilibrium coefficients, respectively. The equilibrium coefficients for the first few moments are [21]:

$$\begin{aligned} \boldsymbol{a}^{(0,eq)} &= \rho \\ \boldsymbol{a}^{(1,eq)} &= \rho \boldsymbol{u} \\ \boldsymbol{a}^{(2,eq)} &= \rho \boldsymbol{u}\boldsymbol{u} + \rho\bar{\theta}\boldsymbol{\delta} \\ \boldsymbol{a}^{(3,eq)} &= \rho \boldsymbol{u}\boldsymbol{u}\boldsymbol{u} + \rho\bar{\theta}[\boldsymbol{u}\boldsymbol{\delta}] \\ \boldsymbol{a}^{(4,eq)} &= \rho \boldsymbol{u}\boldsymbol{u}\boldsymbol{u}\boldsymbol{u} + \rho\bar{\theta}[\boldsymbol{u}^2\boldsymbol{\delta}] - \rho\bar{\theta}^2\boldsymbol{\delta}^2 \end{aligned} \qquad (2.6)$$

To derive the macroscopic equations with the source term, Eq. (2.1) is multiplied by a $m$th-order Hermite polynomial and then summed over the index $i$:

$$\frac{\partial}{\partial t}\sum_i f_i \boldsymbol{\mathcal{H}}^{(m)}(\boldsymbol{\xi}_i) + \frac{\partial}{\partial \boldsymbol{x}} \cdot \sum_i \boldsymbol{\xi}_i f_i \boldsymbol{\mathcal{H}}^{(m)}(\boldsymbol{\xi}_i) = -\frac{1}{\tau}\sum_i f_i^{neq} \boldsymbol{\mathcal{H}}^{(m)}(\boldsymbol{\xi}_i) + \sum_i S_i \boldsymbol{\mathcal{H}}^{(m)}(\boldsymbol{\xi}_i). \qquad (2.7)$$

By introducing $\boldsymbol{a}^{(m)} = \sum_i f_i \boldsymbol{\mathcal{H}}^{(m)}(\boldsymbol{\xi}_i)$ and the relation $\xi_\alpha \boldsymbol{\mathcal{H}}^{(m)}(\boldsymbol{\xi}_i) = \boldsymbol{\mathcal{H}}_\alpha^{(m+1)}(\boldsymbol{\xi}_i) + [\boldsymbol{\delta}_\alpha \boldsymbol{\mathcal{H}}^{(m-1)}(\boldsymbol{\xi}_i)]$, Eq. (2.8) reduces to

$$\frac{\partial}{\partial t}\boldsymbol{a}^{(m)} + \frac{\partial}{\partial x_\alpha} \cdot \left(\boldsymbol{a}^{(m+1)} + [\boldsymbol{\delta}_\alpha \boldsymbol{a}^{(m-1)}]\right) = -\frac{1}{\tau}\boldsymbol{a}^{(m,neq)} + \boldsymbol{s}^{(m)}, \qquad (2.8)$$

The index $\alpha$ appears in Eq. (2.9) means $\alpha$ is excluded from the permutation of indices in the bracket operator. For example, $[\boldsymbol{\delta}\boldsymbol{a}^{(1)}]_{\alpha\beta\gamma} = \delta_{\alpha\beta}a_\gamma^{(1)} + \delta_{\alpha\gamma}a_\beta^{(1)} + \delta_{\beta\gamma}a_\alpha^{(1)}$ contains three terms, while $[\boldsymbol{\delta}_\alpha \boldsymbol{a}^{(1)}]_{\beta\gamma} = \delta_{\alpha\beta}a_\gamma^{(1)} + \delta_{\alpha\gamma}a_\beta^{(1)}$ contains only two terms.

Setting *m*=0, 1, 2, 3 in Eq. (2.8) leads to the conservation equations for different velocity moments:

$$\frac{\partial}{\partial t}a^{(0)} + \frac{\partial}{\partial x_\alpha}a^{(1)} = s^{(0)},$$

$$\frac{\partial}{\partial t}a^{(1)} + \frac{\partial}{\partial x_\alpha}(a^{(2)} + a^{(0)}\delta) = s^{(1)},$$

$$\frac{\partial}{\partial t}a^{(2)} + \frac{\partial}{\partial x_\alpha}(a^{(3)} + [a^{(1)}\delta_\alpha]) = -\frac{1}{\tau}a^{(2,neq)} + s^{(2)}, \quad (2.9)$$

$$\frac{\partial}{\partial t}a^{(3)} + \frac{\partial}{\partial x_\alpha}(a^{(4)} + [a^{(2)}\delta_\alpha]) = -\frac{1}{\tau}a^{(3,neq)} + s^{(3)}$$

After projecting only the coefficient $a^{(n,eq)}$, $a^{(n,neq)}$ and $s^{(n)}$ remain. The non-equilibrium coefficients $a^{(n,neq)}$, modified by the external source $S_i$, can be derived using the Chapman-Enskog expansion similar in Ref. [21]. Details of the derivation refer to Appendix A:

$$a^{(2,neq)} = -\tau\rho\theta\left(\left[\frac{\partial}{\partial x}u\right] - \frac{2}{D}\delta\left(\frac{\partial}{\partial x}\cdot u\right)\right) + \tau\delta([s^{(1)}u] - s^{(2)}):\delta,$$

$$a^{(3,neq)} = -\tau\left[\delta\frac{\partial}{\partial x_\mu}\cdot(\rho u u_\mu)\right] - \tau\rho\theta\left[\frac{\partial}{\partial x}\theta\delta\right] - \tau\rho\theta\frac{\partial}{\partial x}\cdot[u^2\delta] \quad (2.10)$$

$$+\tau\left[\delta\frac{\partial}{\partial x_\epsilon}\rho\theta u_\epsilon u\right] + \tau\rho\theta[\delta u]\frac{2}{D}\nabla\cdot u - \tau\Delta',$$

where

$$\Delta' = -s^{(3)} + [s^{(1)}u^2] + \bar\theta[\delta s^{(1)}] + [u\delta]([-s^{(1)}u + s^{(2)}]:\delta). \quad (2.11)$$

Substituting Eq. (2.10) into Eq. (2.9) yields the macroscopic equations with the source term. The mass equation is written as

$$\frac{\partial}{\partial t}\rho + \frac{\partial}{\partial x_\alpha}\rho u = s^{(0)}. \quad (2.12)$$

The mass equation includes the zeroth and first-order coefficient of $f_i^{eq}$. It is evident that $s^{(0)}$ represents a mass source.

The momentum equation is given by:

$$\frac{\partial}{\partial t}\rho\boldsymbol{u} + \frac{\partial}{\partial \boldsymbol{x}} \cdot \rho\boldsymbol{uu} = \frac{\partial}{\partial \boldsymbol{x}} \cdot (-p\boldsymbol{\delta} + \mathbf{T}) + \boldsymbol{s}^{(1)} - \frac{\partial}{\partial \boldsymbol{x}} \cdot \tau\boldsymbol{\delta}([\boldsymbol{s}^{(1)}\boldsymbol{u}] - \boldsymbol{s}^{(2)})\mathbf{:}\boldsymbol{\delta}, \quad (2.13)$$

where $\boldsymbol{a}^{(2,eq)}$ contributes to the momentum flux tensor $\rho\boldsymbol{uu}$, while $\boldsymbol{s}^{(1)}$ and $\boldsymbol{s}^{(2)}$ contributes to the momentum source. If a momentum source in the continuum form is expressed as a gradient term, then $\boldsymbol{s}^{(1)} = 0$ and $\boldsymbol{s}^{(2)} \neq 0$. Conversely, if the momentum source is unrelated to the gradient of a scalar, then $\boldsymbol{s}^{(1)} \neq 0$ and $\boldsymbol{s}^{(2)} = [\boldsymbol{s}^{(1)}\boldsymbol{u}]$.

The energy equation is obtained by calculating the trace of the stress tensor. It can be written as

$$\begin{aligned}\frac{\partial}{\partial t}(\rho E) + \frac{\partial}{\partial \boldsymbol{x}} \cdot (\rho E\boldsymbol{u}) &= \frac{\partial}{\partial \boldsymbol{x}} \cdot (-p\boldsymbol{\delta}\boldsymbol{u} + \mathbf{T} \cdot \boldsymbol{u}) \\ &+ \frac{D+2}{2}\frac{\partial}{\partial \boldsymbol{x}} \cdot \left(\tau\rho\theta\frac{\partial}{\partial \boldsymbol{x}}\theta\right) \\ &+ \frac{1}{2}\boldsymbol{s}^{(2)}\mathbf{:}\boldsymbol{\delta} + \frac{1}{2}\frac{\partial}{\partial \boldsymbol{x}} \cdot \tau(\boldsymbol{\Delta}'\mathbf{:}\boldsymbol{\delta}) + \text{HOE},\end{aligned} \quad (2.14)$$

where the high-order error term HOE is written as

$$HOE = \frac{D+2}{2}\frac{\partial}{\partial \boldsymbol{x}} \cdot \left\{\tau\boldsymbol{u}^2 \cdot \left(\frac{\partial}{\partial \boldsymbol{x}}\rho\theta\right)\right\} - \frac{D+2}{2}\frac{\partial}{\partial \boldsymbol{x}} \cdot \left\{\tau\frac{\partial}{\partial \boldsymbol{x}} \cdot (\rho\boldsymbol{u}^2)\right\}. \quad (2.15)$$

Eq. (2.13) ~ Eq. (2.15) describes the evolution of mass, momentum, and energy. The high-order error brought by $\boldsymbol{a}^{(3,neq)}$ are typically small unless the fluid exhibits significant compressibility or temperature gradients. The source terms $\boldsymbol{s}^{(1)}$, $\boldsymbol{s}^{(2)}$ and $\boldsymbol{s}^{(3)}$ contribute to the energy equation, with momentum sources directly influencing energy through work done on the fluid. If the energy source involves a gradient term, then $\boldsymbol{s}^{(3)} \neq 0$. If the energy source is independent of the gradient term, then $\boldsymbol{s}^{(2)} \neq 0$ and $\boldsymbol{\Delta}' = 0$. It is important to note that the coefficients of the source term do not strictly correspond to the conservation of specific velocity moments, as both low-order and high-order coefficients can appear within a single equation, particularly in the energy equation.

## 3. Constructing the source term for high-order lattice Boltzmann method

As discussed in the previous section, once the continuum form of the source term is known, it can be systematically constructed. In this section, three typical source terms are derived: the body force source, the thermal diffusion source, and the pressure tensor source. The body force source enables the high-order lattice Boltzmann method to simulate flows driven by body force, the thermal diffusion source allows for the simulation of flows with variable Prandtl numbers. The pressure tensor source facilitates the simulation of multiphase flows with thermodynamic consistency.

### 3.1 The body force

The N-S equation with a body force can be written as

$$\begin{aligned}
&\frac{\partial}{\partial t}\rho + \frac{\partial}{\partial x_\alpha}\rho u = 0 \\
&\frac{\partial}{\partial t}\rho u + \frac{\partial}{\partial x}\cdot \rho u u = \frac{\partial}{\partial x}\cdot(-p\delta + \mathbf{T}) + \mathbf{F} \\
&\frac{\partial}{\partial t}(\rho E) + \frac{\partial}{\partial x}\cdot(\rho E u) = \frac{\partial}{\partial x}\cdot(-p\delta u + \mathbf{T}\cdot u) \\
&\qquad\qquad\qquad\qquad + \frac{D+2}{2}\frac{\partial}{\partial x}\cdot\left(\tau\rho\theta\frac{\partial}{\partial x}\theta\right) + \mathbf{F}\cdot u.
\end{aligned} \qquad (3.1)$$

From the discussion in Section 2.2, the coefficients of $S_i$ are expressed as

$$\begin{aligned}
&s^{(0)} = 0, \quad s^{(n)} = 0 \ (n \geq 4) \\
&s^{(1)} = \mathbf{F}, \\
&s^{(2)} = [s^{(1)} \ \ u] = [\mathbf{F}u], \\
&s^{(3)} = [s^{(1)} \ \ u^2\ ] + \bar{\theta}[us^{(1)} \ \ ] = [\mathbf{F}u^2\ ] + \bar{\theta}[\mathbf{F}u]
\end{aligned} \qquad (3.2)$$

Therefore, the force term $F_i$ can be written as

$$F_i = w_i\{\xi_i \cdot \boldsymbol{F} + (\xi_i \cdot \boldsymbol{u})(\xi_i \cdot \boldsymbol{F}) - (\boldsymbol{u} \cdot \boldsymbol{F})$$
$$+ \frac{1}{2}(\xi_i \cdot \boldsymbol{F})[(\xi_i \cdot \boldsymbol{u})^2 + \bar{\theta}\|\xi_i\|^2 - \|\boldsymbol{u}\|^2 - \bar{\theta}(D+2)] \quad (3.3)$$
$$- (\xi_i \cdot \boldsymbol{u})(\boldsymbol{u} \cdot \boldsymbol{F})\}$$

This body force follows the form presented in Ref [36], and is essentially a He-Luo approximation of the term $(\boldsymbol{F}/\rho) \cdot \nabla f$. Traditionally, the force source is derived in the continuous velocity space. However, it shows that the source can also be derived within the discrete lattice framework.

## 3.2 The thermal diffusion source

Traditional high-order lattice Boltzmann model suffer from a fixed Prandtl number, as the BGK collision operator relaxes all velocity moments at the same rate. The deficiency is usually addressed by introducing a new collision operator. However, such operators tend to be complex due to the need to independently control viscous dissipation and thermal diffusion.

In this section, the thermal diffusion source $H_i$ is introduced to achieve the variable Prandtl number without modifying the collision operator. The source is designed to recover the thermal diffusion term in the energy equation:

$$H_{i,c} = \frac{\partial}{\partial \boldsymbol{x}} \cdot \left(\lambda \frac{\partial}{\partial \boldsymbol{x}} \theta\right), \quad (3.4)$$

where $\lambda = \mu c_p/\mathrm{Pr}$ is the thermal conductivity, Pr is the Prandtl number of the fluid, and $c_p = (D+2)/2$ is the specific thermal capacity. Since the thermal diffusion source affects only the energy equation, only $\boldsymbol{s}^{(3)}$ is non-zero. Thus,

$$\frac{1}{2}\frac{\partial}{\partial \boldsymbol{x}} \cdot \tau(-\boldsymbol{s}^{(3)} : \boldsymbol{\delta}) = \frac{D+2}{2}\frac{\partial}{\partial \boldsymbol{x}} \cdot \left(\frac{\tau\rho\theta}{\mathrm{Pr}} \frac{\partial}{\partial \boldsymbol{x}}\theta\right) - \frac{D+2}{2}\frac{\partial}{\partial \boldsymbol{x}} \cdot \left(\tau\rho\theta \frac{\partial}{\partial \boldsymbol{x}}\theta\right) \quad (3.5)$$

and

$$s^{(3)} = \rho\theta\left(1 - \frac{1}{\Pr}\right)\left[\boldsymbol{\delta}\frac{\partial}{\partial \boldsymbol{x}}\theta\right] \tag{3.6}$$

The heat flux source $H_i$ can be written as

$$\begin{aligned} H_i &= \frac{1}{6}w_i\rho\theta\left(1 - \frac{1}{\Pr}\right)\left[\boldsymbol{\delta}\frac{\partial}{\partial \boldsymbol{x}}\theta\right]\boldsymbol{\mathcal{H}}^{(3)}(\boldsymbol{\xi}_i) \\ &= \frac{1}{2}w_i\rho\theta\left(1 - \frac{1}{\Pr}\right)(\|\boldsymbol{\xi}_i\|^2 - D - 2)\boldsymbol{\xi}_i \cdot \frac{\partial}{\partial \boldsymbol{x}}\theta \end{aligned} \tag{3.7}$$

The temperature gradient can be calculated by the finite difference method. By adjusting the thermodynamic conductivity, the Prandtl number can be varied, ensuring that the heat transfer process remains consistent with the actual physical phenomena.

## 3.3 The pressure tensor source

The pseudopotential LBM is widely applied to simulate multiphase flows [37]. In this section, the pseudopotential LBM is extended to the high-order lattice. The pseudopotential LBM incorporates the capillary tensor via a pseudopotential force:

$$\boldsymbol{F}^{(SC)} = -G\psi(\boldsymbol{x})\sum_i w_i^{(SC)}\psi(\boldsymbol{x} + \boldsymbol{\xi}_i\Delta t)\boldsymbol{\xi}_i\Delta t, \tag{3.8}$$

with $G = -1$ and $\psi(\rho) = \sqrt{2(p_{EOS}(\rho) - \rho\theta)/G}$. $p_{EOS}$ is the equation of state. If the pseudopotential force is introduced by the method proposed in Section 3.1, the capillary tensor $\mathbf{C}$ of the pseudopotential model can be written as [38]:

$$\mathbf{C} = -\frac{G}{3}\left(\left(\frac{a}{2}(\nabla\psi)\cdot(\nabla\psi) + d\psi\nabla\cdot\nabla\psi\right)\boldsymbol{\delta} + c\nabla\psi\otimes\nabla\psi + b\psi\nabla\nabla\psi\right) \tag{3.9}$$

The coefficients $a$, $b$, $c$, and $d$ are undetermined, but their combination $\epsilon = -(2a + c)/(b + d)$ is kept to zero. This leads to a significant underestimation of gas density [37], requiring an additional source term to accurately model liquid-gas densities.

The continuum form of the pressure tensor source is written as

$$P_{i,c} = -Gk\nabla \cdot \big((\nabla\psi) \cdot (\nabla\psi)\big)\boldsymbol{\delta} \tag{3.10}$$

where $k$ is a free parameter that adjusts $\epsilon$. Since $P_{i,c}$ acts as a momentum source, only $s^{(2)}$ is non-zero:

$$\frac{\partial}{\partial \boldsymbol{x}} \cdot \big(\tau\boldsymbol{\delta}\ \text{tr}(s^{(2)})\big) = -Gk\nabla \cdot \big((\nabla\psi) \cdot (\nabla\psi)\big)\boldsymbol{\delta}. \tag{3.11}$$

Therefore, the pressure tensor source is written as

$$P_i = -w_i \frac{k}{2D\tau} \frac{\|\boldsymbol{F}^{(SC)}\|^2}{\psi^2 G}\ (\|\boldsymbol{\xi}_i\|^2 - D), \tag{3.12}$$

with $(\nabla\psi) \cdot (\nabla\psi) \approx \|\boldsymbol{F}^{(SC)}\|^2/(G^2\psi^2)$. By adjusting $k$, the liquid-gas density can be controlled. The optimal value of $k$ depends on a specific equation of state. The momentum sources for multiphase flows are composed of two source terms, $F_i^{SC}$ and $P_i$, where $F_i^{SC}$ introduces the pseudopotential force, and $P_i$ adjusts the liquid-gas density.

## 4. Implementation of the high-order lattice Boltzmann method

The previous section presented the standard method for constructing the body source. However, when applying this method to a specific high-order lattice, certain modifications must be made. In this section, the key modifications are introduced, with the multispeed LBM serving as an example. The modifications for other high-order lattices can be made analogously.

### 4.1 Scaling of the discrete velocity

In this study, the D2Q37 lattice is employed [14]. It provides 37 discrete velocities and eighth-order isotropy. The discrete velocities are shown in Fig. 1, and the gas constant

$c_0 \approx 0.83543$. The corresponding weights can be found in Table 1.

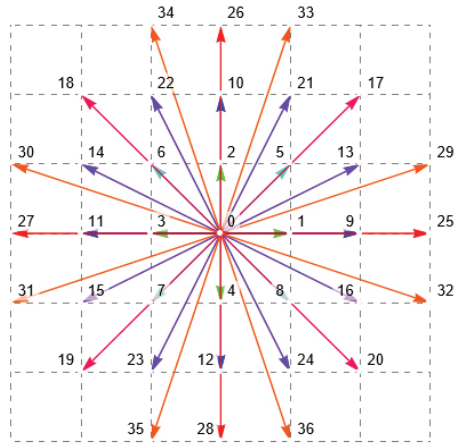

Fig. 1 The discrete velocities of the D2Q37 lattice.

Table 1 Weights of the D2Q37 lattice

| $i$ | $w_i$ |
| --- | --- |
| 0 | 0.233151 |
| 1~4 | 0.107306 |
| 5~8 | 0.0576679 |
| 9~12 | 0.0142082 |
| 13~16 | 0.00535305 |
| 17~20 | 0.00101194 |
| 21~24 | 0.00535305 |
| 25~28 | 0.000245301 |
| 29~36 | 0.000283414 |

Compared to the D2Q9 lattice, the D2Q37 provides eighth-order isotropy, making it suited for simulating thermal flows. Another advantage of the D2Q37 lattice is the lattice constant $c_0$ is larger than D2Q9, reducing the local compressibility of the fluid and thus expanding the feasible region for local velocities in the positive equilibrium distribution function.

The discrete velocities $c_i$ of D2Q37 lattice are related to the integral points $\xi_i$ via the lattice constant, namely $c_i = \xi_i/c_0$. Consequently, the formulation presented in Section 2 must be scaled appropriately. The scaled terms are summarized in Table 2.

Table 2. Major terms after the scaling of discrete velocity

| Name | Form |
|---|---|
| $f_i^{eq}$ | $\begin{aligned} f_i^{eq} = w_i \rho \Big\{ & 1 + \frac{1}{c_0^2} c_i \cdot u + \frac{1}{2c_0^4}(c_i \cdot u)^2 - \frac{1}{2c_0^2}\|u\|^2 + \frac{1}{2c_0^2}\bar{\theta}(\|c_i\|^2 - c_0^2 D) \\ & + \frac{1}{6c_0^6}(c_i \cdot u)^3 - \frac{1}{2c_0^4}(c_i \cdot u)\|u\|^2 + \frac{1}{2c_0^4}(c_i \cdot u)\bar{\theta}[\|c_i\|^2 - c_0^2(D \\ & + \frac{1}{24c_0^8}(c_i \cdot u)^4 - \frac{1}{4c_0^6}(c_i \cdot u)^2\|u\|^2 + \frac{1}{8c_0^4}\|u\|^4 \\ & + \frac{1}{4c_0^6}\bar{\theta}\big[(c_i \cdot u)^2(\|c_i\|^2 - c_0^2(D+4)) + c_0^2\|u\|^2(c_0^2(D+2) - \|c \\ & + \frac{1}{8c_0^4}\bar{\theta}^2\big[\|c_i\|^4 - 2c_0^2(D+2)\|c_i\|^2 + c_0^4 D(D+2)\big] \Big\} \end{aligned}$ |
| $F_i$ | $\begin{aligned} F_i^{3rd} = w_i \Big\{ & \frac{1}{c_0^2} c_i \cdot F + \frac{1}{c_0^4}(c_i \cdot u)(c_i \cdot F) - \frac{1}{c_0^2}(u \cdot F) \\ & + \frac{1}{2c_0^6}(c_i \cdot F)\big[(c_i \cdot u)^2 + \bar{\theta}c_0^2\|c_i\|^2 - c_0^2\|u\|^2 - \bar{\theta}c_0^4(D+2)\big] \\ & - \frac{1}{c_0^4}(c_i \cdot u)(c_i \cdot F) \Big\} \end{aligned}$ |
| $H_i$ | $H_i = \frac{1}{2c_0^2} w_i \rho \theta \left(1 - \frac{1}{c_0^2 \text{Pr}}\right)(\|c_i\|^2 - c_0^2(D+2))c_i \cdot \frac{\partial}{\partial x}\theta$ |
| $P_i$ | $P_i = -w_i \frac{k}{2D\tau} \frac{\|F^{(SC)}\|^2}{\psi^2 G}(\|c_i\|^2 - c_0^2 D)$ |

## 4.2 Spatial and temporal discretization

The spatial and temporal discretization follows the same procedure as that in the isothermal lattice Boltzmann model. The lattice Boltzmann equation can be integrated along the characteristics, i.e., $s(\zeta): x = x(\zeta), t = t(\zeta)$, where $\zeta$ is the parametric variable. For simplicity the characteristic line $s(t): x = c_i t, t = t$ is chosen. The derivative of the distribution function for $t$ is then expressed as

$$\frac{d}{dt}f = \frac{\partial}{\partial t}\frac{\partial t}{\partial \zeta}f + \frac{\partial}{\partial x} \cdot c_i f \ . \tag{4.1}$$

Integrating Eq. (2.1) yields

$$\int_t^{t+\Delta t} \frac{d}{dt} f_i d\zeta = \int_t^{t+\Delta t} (\Omega_i + S_i) d\zeta \\ = \frac{1}{2}\Delta t\{[\Omega_i(t+\Delta t) + S_i(t+\Delta t)] + [\Omega_i(t) + S_i(t)]\}, \tag{4.2}$$

where a trapezoidal integration is adopted for the right-hand side. To avoid implicit iteration of time a "shift" operation is performed:

$$\hat{f}_i = f_i - \frac{\Delta t}{2}(\Omega_i + S_i), \\ \hat{\tau} = \tau + \frac{\Delta t}{2}. \tag{4.3}$$

Substituting Eq. (4.3) into Eq. (4.2) yields

$$\hat{f}_i(x + c_i \Delta t, t + \Delta t) - \hat{f}_i(x, t) = -\frac{\Delta t}{\hat{\tau}}(\hat{f}_i - f_i^{eq}) + \left(1 - \frac{\Delta t}{2\hat{\tau}}\right) S_i \Delta t, \tag{4.4}$$

Although Eq. (4.4) is derived originally for isothermal LBM, it is also suited for the high-order LBM. By defining the modified velocity moment $\hat{\mathcal{F}}_k = \sum_i c_i^k \hat{f}_i$ and $\mathcal{G}_k = \sum_i c_i^k S_i$, the macroscopic parameters are calculated in terms of $\hat{\mathcal{F}}_k$ and $\mathcal{G}_k$:

$$\rho = \hat{\mathcal{F}}_0, \\ u = \frac{1}{\rho}\hat{\mathcal{F}}_1 + \frac{\Delta t}{2}\mathcal{G}_1, \\ \theta = \frac{1}{\rho D c_0^2}(\hat{\mathcal{F}}_2 \cdot \boldsymbol{\delta} - \rho\|u\|^2) + \frac{\Delta t}{2\rho D c_0^2}\mathcal{G}_2 \cdot \boldsymbol{\delta}. \tag{4.5}$$

The non-equilibrium distribution function is given by

$$\hat{f}_i^{neq} = f_i^{neq} - \frac{\Delta t}{2}(\Omega_i + S_i) = \frac{2\hat{\tau}}{2\hat{\tau} - \Delta t} f_i^{neq} - \frac{\Delta t}{2} S_i \tag{4.6}$$

Eq. (4.6) highlights the role of the source term $S_i$ in high-order LBM. It acts as a non-equilibrium effect of the distribution function. During the iterations, the relaxation of the source term is also excluded. The physical effects brought by the source term are not realized through kinetic theory, but rather by exploiting the mathematical properties of the equations, directly incorporating the expected macroscopic effects into the iteration process. This method, while lacking a clear physical interpretation, offers practical advantages by bypassing the need to delve into the complex mesoscopic flow mechanisms. This significantly simplifies the implementation process, making the method both efficient and straightforward for practical use.

## 4.3 The boundary conditions

The treatment of boundary conditions in high-order LBM differs significantly from that in the standard LBM. In the BGK scheme, boundary nodes interact only with the nearest lattice layer. However, in high-order lattice models, discrete velocities span multiple lattice nodes (Fig. 2), which introduces additional complexity. The boundary treatment must account for these extended interactions, resulting in more unknown distribution functions and boundary nodes.

To address these challenges, various boundary conditions have been developed. For open boundaries, characteristic boundary methods are commonly used [39]. For wall boundaries, several strategies are available, including diffuse boundary conditions [26], regularized boundaries [40], and the Tamm-Mott-Smith (TMS) boundaries [22]. In this study, the TMS boundary condition is used. In the TMS method, the boundary nodes are categorized as wall nodes and fluid nodes (Fig. 2). The unknown distribution functions at the boundary are approximated as a weighted combination of distribution functions from two distinct states:

$$f_i = f_{\bar{\imath}} + f_i^{eq}(\rho_{tgt}, \boldsymbol{u}_{tgt}, T_{tgt}) - f_i^{eq}(\rho_{loc}, \boldsymbol{u}_{loc}, T_{loc}), \tag{4.7}$$

where $f_{\bar{\imath}}$ is the distribution function in the opposite direction. The target macroscopic quantities are calculated based on the boundary condition for wall nodes and are kept as the local macroscopic quantities in the previous time step for fluid nodes. The local macroscopic quantities are calculated as follows:

$$\begin{aligned} \rho_{loc} &= \sum_{i \in \mathcal{U}} f_i^{eq}(\rho_{tgt}, \boldsymbol{u}_{tgt}, T_{tgt}) + \sum_{i \in \mathcal{K}} f_i, \\ \rho_{loc} \boldsymbol{u}_{loc} &= \sum_{i \in \mathcal{U}} f_i^{eq}(\rho_{tgt}, \boldsymbol{u}_{tgt}, T_{tgt}) \boldsymbol{c}_i + \sum_{i \in \mathcal{K}} f_i \boldsymbol{c}_i + \frac{\Delta t}{2} \mathcal{G}_1. \end{aligned} \tag{4.8}$$

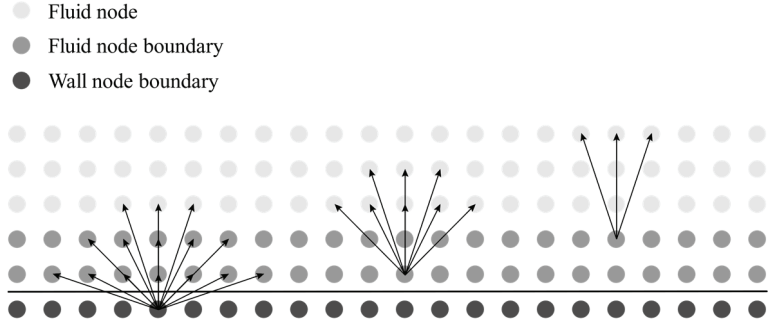

Fig. 2 The boundary nodes of the TMS boundary. The solid line represents the solid-liquid interface.

## 5. Numerical results and discussion

In this section, several numerical tests are performed to validate the source terms under various scenarios. At first, thermal Couette flow and thermal Poiseuille flow are examined to validate the heat dissipation source. Next, the Rayleigh-Bénard convection, where the fluid density is influenced by both temperature and body force, is simulated. Finally, the liquid-gas coexistence curve and spurious velocity are discussed in the context of the pressure tensor source.

## 5.1 The thermal Couette flow

The schematic of the problem is shown in Fig. 3. The fluid between two infinite parallel plates is driven by the motion of the top wall with a constant velocity $u_0$.

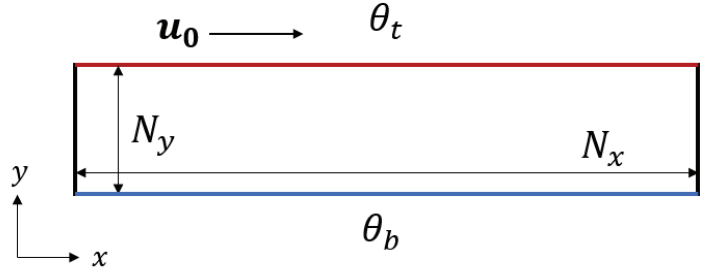

Fig. 3 The schematic of the thermal Couette flow.

If $\theta_t > \theta_b$, the velocity and the non-dimensional temperature $\theta^* = (\theta - \theta_b)/(\theta_t - \theta_b)$ along the y-axis are given by [41]

$$u_x = u_0 y^*, u_y = 0, \qquad (5.1)$$

$$\theta^* = y^* + \frac{1}{2} \Pr \cdot \text{Ec } y^*(1 - y^*), \qquad (5.2)$$

where $y^* = y/H$ is the non-dimensional y coordinate, Pr is the Prandtl number, and Ec is the Eckert number:

$$\text{Ec} = \frac{\text{Re}^2 \nu^2}{H^2 c_p (\theta_h - \theta_c)} \qquad (5.3)$$

The simulation is conducted on a 50×56 grid, with the lattice viscosity $\nu = c_0^2(\hat{\tau} - 0.5)$ set to 0.01 and Re=100. Ec ranges from 1 to 20, and Pr ranges from 0.71 to 1.41. The computational domain is periodic along the x-axis, with wall boundaries applied to the top and bottom walls. For the TMS boundary, the actual height of the domain is 54. The fluid is initialized with zero velocity and a constant temperature $\theta_t = 0.8$.

To simulate flow under a wide range of Pr and Ec, the heat dissipation source $H_i$ is included. The velocity profile along the y-axis is presented in Fig. 4, and the dimensionless temperature profiles along the y-axis are shown in Fig. 5. The solid lines in both figures represent the theoretical predictions. The results have good agreement with theoretical prediction, confirming that $H_i$ accurately captures the heat dissipation process in the flow.

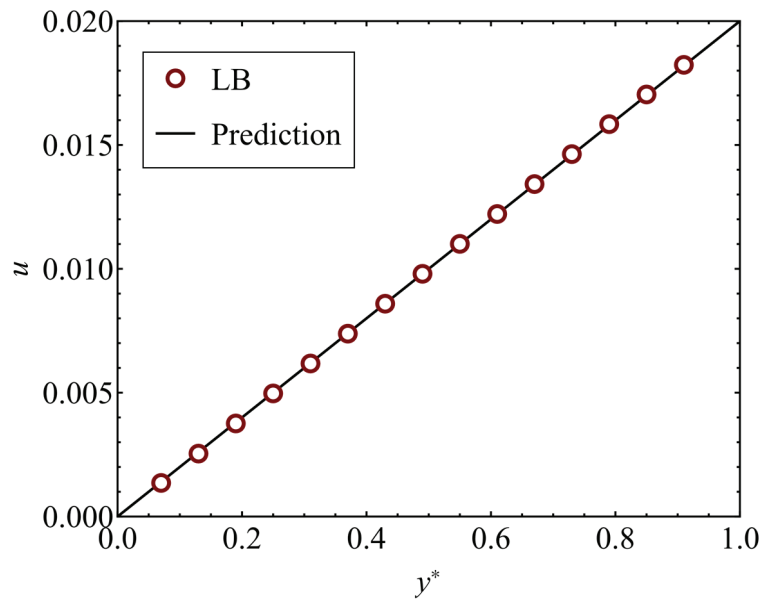

Fig. 4 Velocity along the center line.

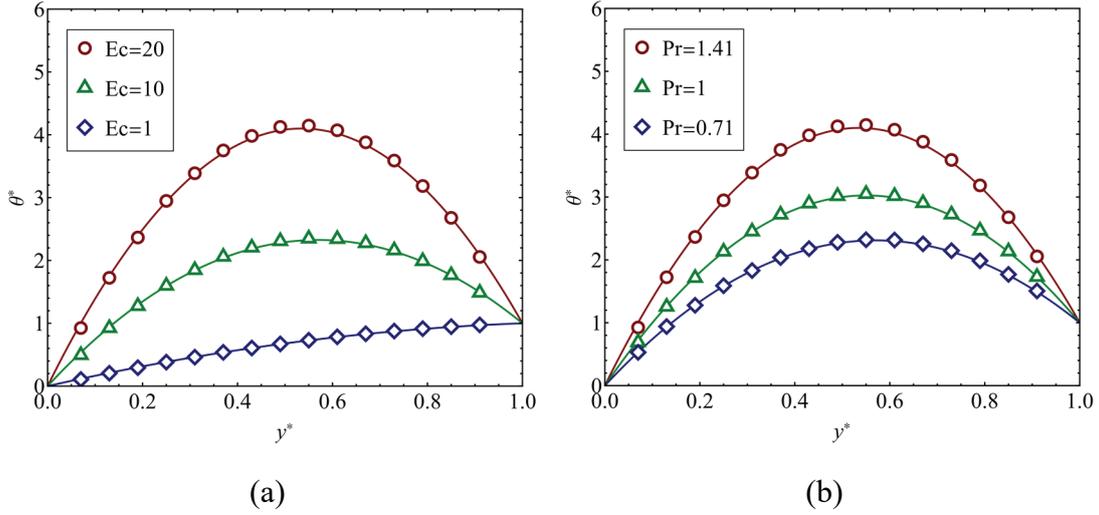

Fig. 5 Dimensionless temperature along the *y*-axis with various Pr and Ec.

## 5.2 The thermal Poiseuille flow

The schematic of the problem is shown in Fig. 6. The fluid between two infinite horizontal plates is driven by a horizontal acceleration $g$. The distance between two plates is $H$, with the temperature of the top wall set to $\theta_t$ and the bottom wall to $\theta_b$. If $\theta_t > \theta_b$, the velocity and the non-dimensional temperature $\theta^* = (\theta - \theta_b)/(\theta_t - \theta_b)$ along the y-axis are given by [41]

$$u_x = \frac{gH^2}{2\rho_0\theta_0\nu} y^*(1-y^*), u_y = 0, \qquad (5.4)$$

$$\theta^* = y^* + \frac{1}{3}\text{Pr}\cdot\text{Ec}\ (1-(1-2y^*)^4), \qquad (5.5)$$

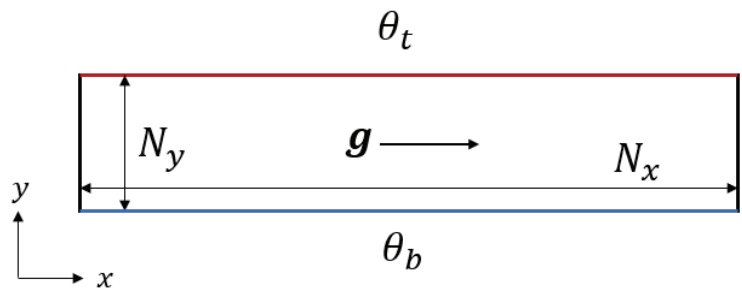

Fig. 6 The schematic of the thermal Poiseuille flow.

The simulation is conducted on a 50×56 grid, for the TMS boundary the actual height

$H = 54$. The viscosity $\nu$ is set to 0.01. The Reynolds number Re is kept as 100, Ec ranges from 1 to 20, and Pr ranges from 0.71 to 1.41. Both the heat dissipation source $H_i$ and the body force source $F_i$ are incorporated into the simulation. The velocity and temperature profiles along the y-axis are shown in Fig. 7 and Fig. 8. The solid lines represent the exact solution based on Eq. (5.4) and Eq. (5.5), while the open markers represent the results obtained from the high-order LBM. The numerical results are in close agreement with the analytical solutions.

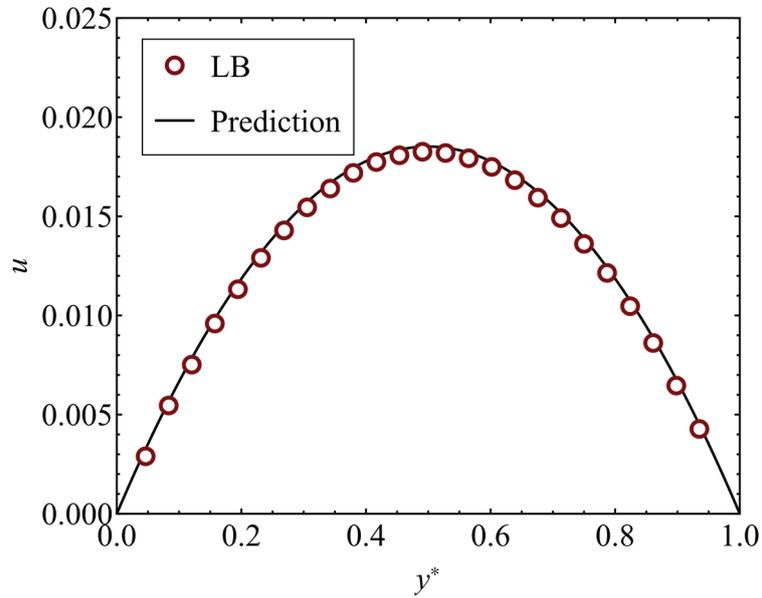

Fig. 7 The velocity profile along the $y$-axis.

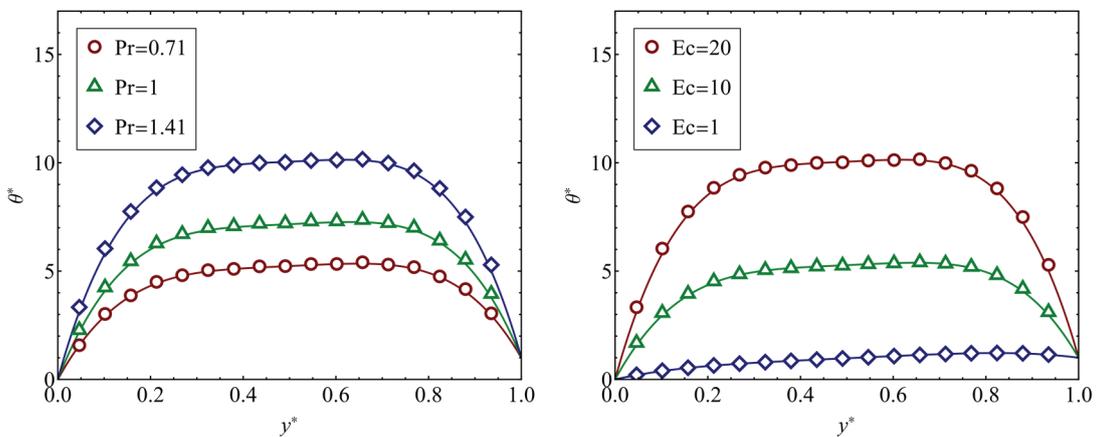

(a) $Ec = 20$  (b) $Pr = 0.71$

Fig. 8 Dimensionless temperature profile along the *y*-axis.

## 5.3 The Rayleigh-Bénard convection

Rayleigh-Bénard convection serves as a classical benchmark for thermal flows. In this case, a horizontal fluid layer is heated from below, while the upper surface is kept at a cooler, constant temperature. This temperature difference induces a buoyancy-driven flow, where the warmer fluid at the bottom becomes less dense and rises, while the cooler fluid near the top sinks, forming convective cells. This process is governed by the interplay between thermal diffusion and buoyancy forces, with the dimensionless Rayleigh number (Ra) serving as a critical parameter that dictates whether convection occurs. The schematic of the problem is shown in Fig. 9.

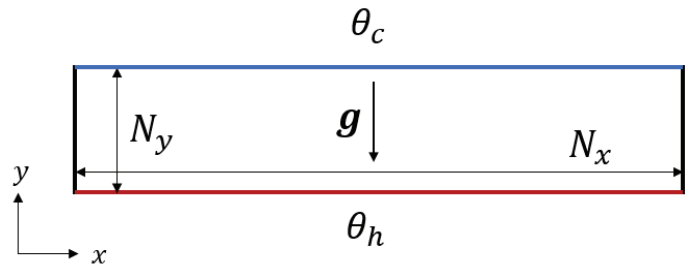

Fig. 9 The schematic of Rayleigh- Bénard convection.

The simulation is conducted on a 200×102 grid, namely the domain length $L = 200$ and height $H = 100$. The Rayleigh number is defined as follows:

$$\mathrm{Ra} = 2\mathrm{Pr}\frac{g(\theta_h - \theta_c)H^3}{\nu^2(\theta_h + \theta_c)}, \tag{5.6}$$

where $\nu = 0.04$, $\theta_c = 1.0$ and $\theta_h = 1.4$. The Prandtl number Pr is fixed at 1.41, and different Rayleigh numbers are obtained by adjusting the gravity $g$. In this study,

Ra ranges from 2000 to 30000. For high Ra flows a high-order lattice with better stability is required, which needs further investigation.

The body force $F_i$ and the heat dissipation source $H_i$ are included in the simulation. The initial temperature is defined as the sum of a linear function of the $y$ coordinate and a sinuous fluctuation:

$$\theta_{init} = \theta_t - (\theta_t - \theta_b)\frac{y}{H} + 0.01(\theta_h - \theta_c)\sin\left(\frac{\pi x}{L}\right)\sin\left(\frac{\pi y}{H}\right) \quad (5.7)$$

Temperature contours for different Ra values are shown in Fig. 10, with ten equally spaced temperature contours labeled as gray lines. It is observed that the heat transfer process is dominated by conduction for small Ra numbers and by convection for large Ra. The Nusselt number is calculated for quantitative analysis:

$$Nu = -\frac{H}{L(\theta_h - \theta_c)}\sum_{k=0}^{L-1}\left(\frac{\partial \theta}{\partial y}\right)_k, \quad (5.8)$$

where $(\partial \theta/\partial y)_k$ is the temperature gradient at $x = k$. The temperature gradient is calculated by a second-order finite difference method.

In Fig. 11 the average Nusselt number for the hot and cold walls is compared with the results obtained from Ref [23] and the empirical formula $Nu = 1.56(Ra/Ra_c)^{0.296}$ [42]. The results show that current model gives the accurate prediction at the moderate Rayleigh numbers.

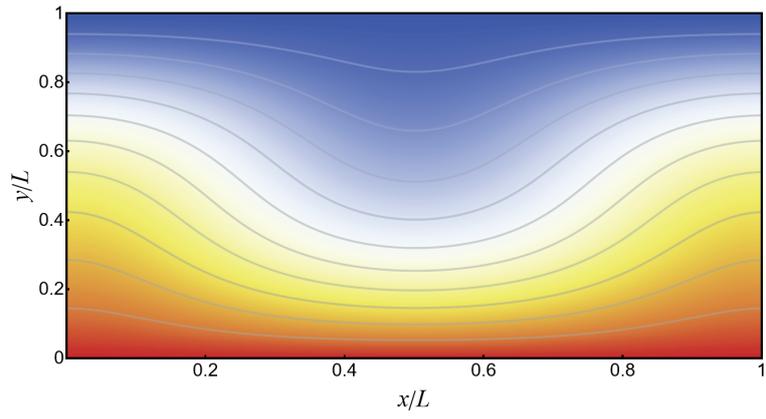

(a) Ra=2500

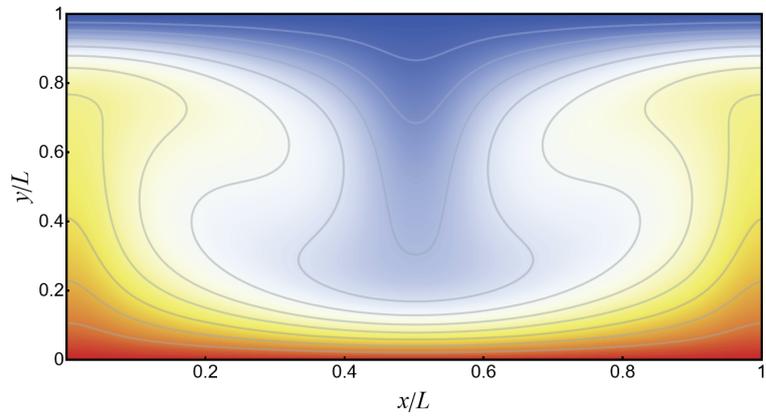

(b) Ra=20000

Fig. 10 Temperature contours of the Rayleigh-Bénard convection.

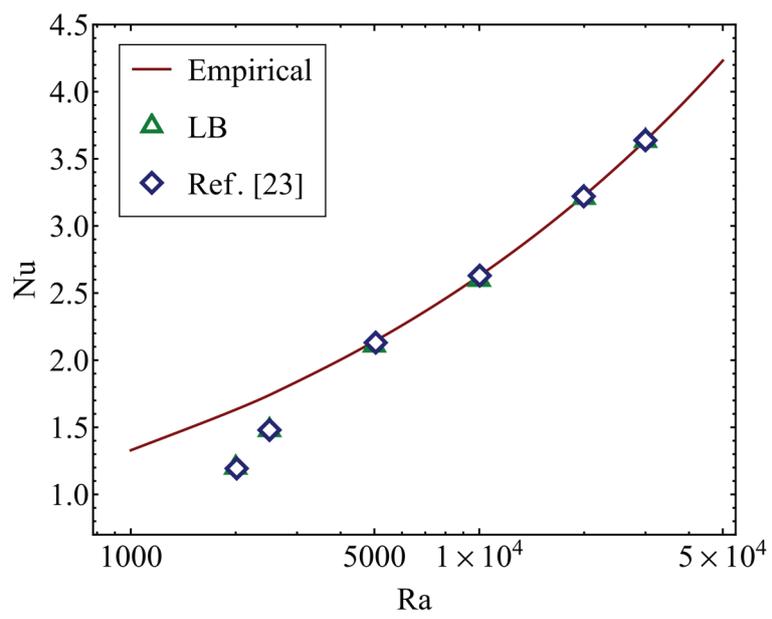

Fig. 11 The average Nusselt number of Rayleigh-Bénard convection

## 5.4 Multiphase flows

The high-order LBM can be also applied to multiphase flows when the pseudopotential force $F_i^{SC}$ and the pressure tensor source $P_i$ are included. For multiphase flows the equation of state which allows for a liquid-gas coexistence state is needed. In this study, the Carnahan-Starling (C-S) equation of state is used:

$$p_{EOS} = \rho R_g T \frac{1 + b\rho/4 + (b\rho/4)^2 - (b\rho/4)^3}{(1 - b\rho/4)^3} - a\rho^2, \qquad (5.9)$$

where $a = 0.25$, $b = 4$ and $R_g = 1$ [37]. Therefore, the critical temperature $T_c \approx 0.023$.

In this section, the high-order lattice Boltzmann method (LBM) is used exclusively to solve the mass and momentum equations, with the temperature $\theta$ kept as one. This choice is dictated by the fact that the temperature parameter in the C-S EOS is significantly lower than one, which would result in a negative equilibrium distribution function. Consequently, the temperature in the C-S EOS serves solely as a parameter influencing the liquid-gas density, without affecting the velocity moments of the distribution function. For clarity, this temperature is denoted as $T$. When dealing with multiphase flows that involve heat transfer, the double distribution function method can be applied, in which the parameter $T$ is evolved separately.

The equilibrium liquid-gas density is simulated on a 50×200 grid with $\nu = 0.2$. The density of the field is initialized as:

$$\rho(y) = \rho_{g,0} + \frac{\rho_{l,0} - \rho_{g,0}}{2} \times \text{abs}\left\{\tanh\left[\frac{2(y-50)}{W}\right] - \tanh\left[\frac{2(y-150)}{W}\right]\right\}, \quad (5.10)$$

where $W = 5$ is the interface width. The liquid domain is located at the center of the grid with a height of 100, and the remainder of the domain is filled with gas. $\rho_{l,0}$ and $\rho_{g,0}$ are equilibrium densities calculated by the Maxwell construction rule. Ideally, when the simulation reaches a steady state, the liquid-gas densities should remain unchanged. It can be achieved by appropriately setting the parameter $k$ in the pressure tensor source $P_i$. For the C-S EOS, the optimal $k$ is 3.02. The value may vary for other EOS. The equilibrium liquid-gas densities, as a function of the reduced temperature $T_r = T/T_c$ are shown in Fig. 12. The solid line denotes the equilibrium densities $\rho_{l,0}$ and $\rho_{g,0}$ for different $T_r$. The two results have a good agreement.

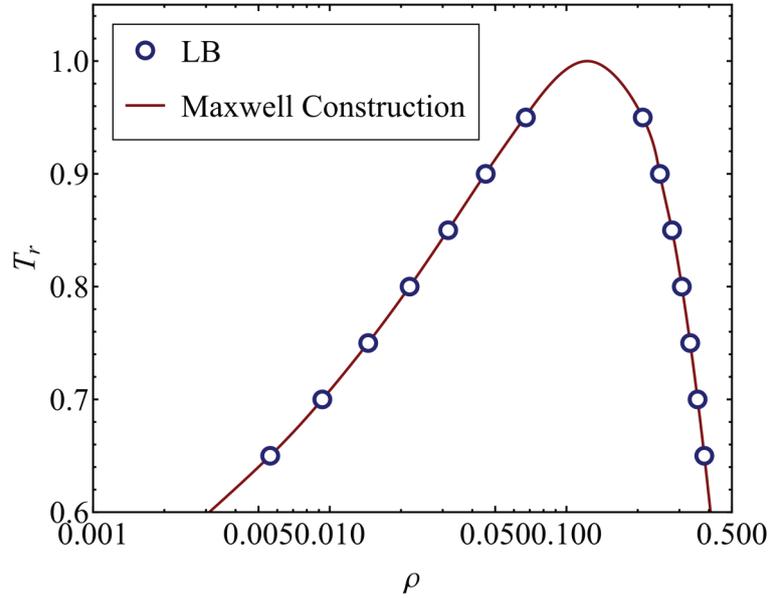

Fig. 12 The equilibrium liquid-gas density.

The spurious velocity is a common issue in pseudopotential models [15]. This results from the pseudopotential force disrupting local momentum conservation, and from insufficient isotropy of traditional isothermal lattices. This leads to spurious currents near the liquid-gas interface. In the high-order LBM, spurious velocities can be significantly reduced due to the enhanced lattice isotropy and the broader distribution of the pseudopotential force across more lattice nodes.

To illustrate this, a droplet located at the center of a 200×200 grid is simulated, with a radius $r_0 = 50$. The reduced temperature $T_r$ ranges from 0.7 to 0.95. The viscosity ν is set to 0.2. Periodic boundaries are applied to all sides of the domain. The density field is initialized as

$$\rho(x,y) = \begin{cases} \dfrac{\rho_{l,0} + \rho_{g,0}}{2} - \dfrac{\rho_{l,0} - \rho_{g,0}}{2} \times \tanh\left[\dfrac{2(d - r_0)}{W}\right], & \text{droplets,} \\ \dfrac{\rho_{l,0} + \rho_{g,0}}{2} + \dfrac{\rho_{l,0} - \rho_{g,0}}{2} \times \tanh\left[\dfrac{2(d - r_0)}{W}\right], & \text{bubbles,} \end{cases} \quad (5.11)$$

where $d$ is the distance from the center of the domain. The maximum spurious velocity of the steady-state solution $\|\boldsymbol{u}\|_{\max}$ as a function of the reduced temperature $T_r = T/T_c$ is shown in Fig. 13. A velocity contour of the droplet under $T_r = 0.7$ is presented in Fig. 14. The results demonstrate that the maximum spurious velocity is two orders of magnitude lower than that in the D2Q9 model, demonstrating the effectiveness of the high-order LBM in suppressing spurious velocities.

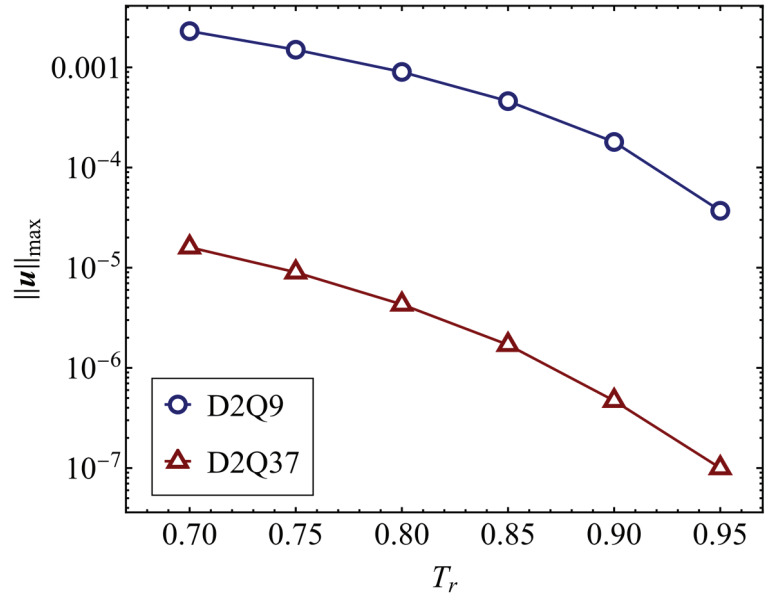

Fig. 13 The maximum spurious velocity of the droplet.

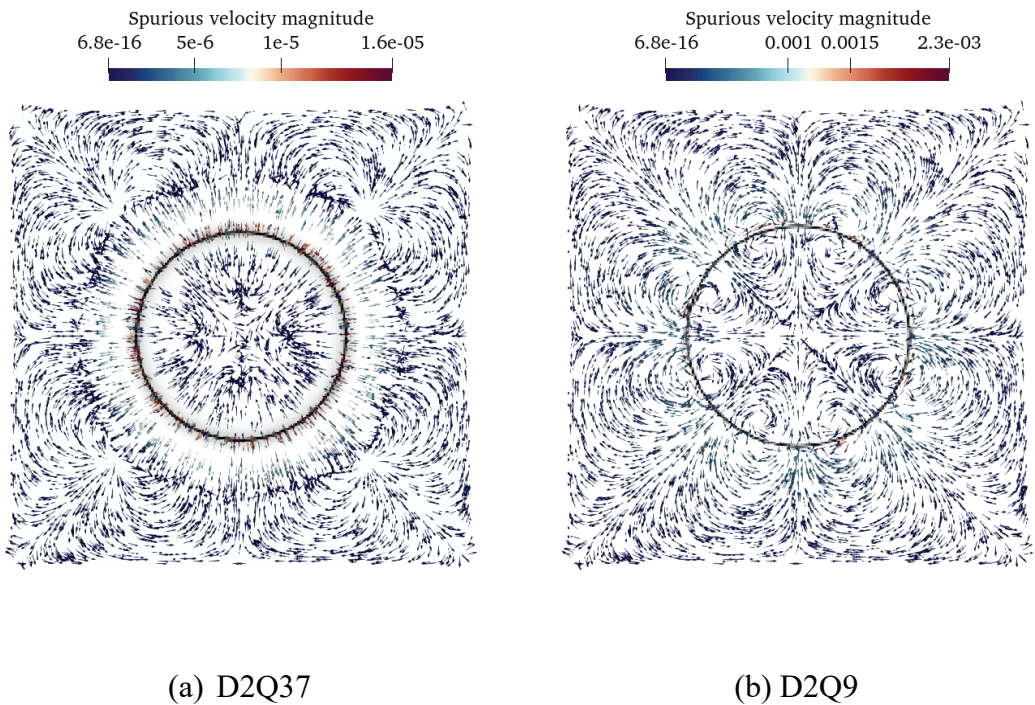

(a) D2Q37  (b) D2Q9

Fig. 14 The spurious currents of the droplet.

These results indicate that the high-order lattice Boltzmann method, with the inclusion of the pseudopotential force and pressure tensor source, can accurately simulate multiphase flows and effectively suppress spurious velocities. Although only static cases are considered here, this method can be readily applied to dynamic scenarios once the initial conditions and driving forces are specified.

## 6. Conclusion

In this paper, a generalized procedure for constructing source terms is presented for the high-order lattice Boltzmann method. The Chapman-Enskog analysis proves that the source term retrieves the expected momentum and energy conservation equation. The flexibility of this approach is highlighted by its ability to construct source terms for a variety of applications. Notably, the method simplifies implementation by bypassing the need to directly control the behavior of high-order moments in the collision operator. Numerical results confirm both the versatility and accuracy of the source term, showcasing its effectiveness in simulating complex flow phenomena. While the multispeed LBM is used in this study for illustration, the method is inherently adaptable and can be applied to other high-order lattices. This provides a robust and efficient framework for simulating a wide range of fluid dynamics problems.

## Appendix A The Chapman-Enskog expansion of the multispeed lattice Boltzmann method

The Chapman-Enskog expansion assumes that the non-equilibrium distribution function $f_i^{neq}$ can be expressed as $\varepsilon f_i^{(1)}$, where $\varepsilon$ is a small quantity related to the Knudsen number. Meanwhile, the time and position are expressed as $t = t(t_1, t_2)$ and $x = x(x_1)$, where $t_1 = \varepsilon t, t_2 = \varepsilon^2 t$ and $x_1 = \varepsilon x$. The time and spatial derivatives are expressed as $\partial/\partial t = \varepsilon \partial/\partial t_1 + \varepsilon^2 \partial/\partial t_2$ and $\partial/\partial x = \varepsilon \partial/\partial x_1$ respectively.

Substituting these multi-scale variables into Eq. (2.1) the results can be re-arranged and split based on the order of $\varepsilon$. The equation of the order $O(\varepsilon)$ is written as

$$O(\varepsilon): \frac{\partial}{\partial t_1} f_i^{(0)} + \frac{\partial}{\partial x_1} \cdot \xi_i f_i^{(0)} = -\frac{1}{\tau} f_i^{(1)} + F_i^{(1)}. \tag{A1}$$

The equation of the order $O(\varepsilon^2)$ is written as

$$O(\varepsilon^2): \frac{\partial}{\partial t_2} f_i^{(0)} + \frac{\partial}{\partial x_1} \cdot \xi_i f_i^{(1)} = 0. \tag{A2}$$

$f_i^{(1)}$ is expressed as

$$f_i^{(1)} = w_i \sum_{n=0}^{\infty} \frac{1}{n!} \boldsymbol{a}^{(n,1)} \cdot \boldsymbol{\mathcal{H}}^{(n)}(\xi_i), i = 1,2,\cdots,N. \tag{A3}$$

$f_i^{(1)}$ is accessible once $\boldsymbol{a}^{(n,1)}$ is identified. Therefore, the same projection operation is conducted on Eq. (A1), which leads to

$$\begin{aligned}
\frac{\partial}{\partial t_1} \boldsymbol{a}^{(0,eq)} + \frac{\partial}{\partial x_{1\alpha}} \boldsymbol{a}^{(1,eq)} &= -\frac{1}{\tau} \boldsymbol{a}^{(0,1)}, \\
\frac{\partial}{\partial t_1} \boldsymbol{a}^{(1,eq)} + \frac{\partial}{\partial x_{1\alpha}} \left( \boldsymbol{a}^{(2,eq)} + \boldsymbol{a}^{(0,eq)} \boldsymbol{\delta} \right) &= -\frac{1}{\tau} \boldsymbol{a}^{(1,1)} + \boldsymbol{s}^{(1)}, \\
\frac{\partial}{\partial t_1} \boldsymbol{a}^{(2,eq)} + \frac{\partial}{\partial x_{1\alpha}} \left( \boldsymbol{a}^{(3,eq)} + [\boldsymbol{a}^{(1,eq)} \boldsymbol{\delta}_\alpha] \right) &= -\frac{1}{\tau} \boldsymbol{a}^{(2,1)} + \boldsymbol{s}^{(2)}, \\
\frac{\partial}{\partial t_1} \boldsymbol{a}^{(3,eq)} + \frac{\partial}{\partial x_{1\alpha}} \left( \boldsymbol{a}^{(4,eq)} + [\boldsymbol{a}^{(2,eq)} \boldsymbol{\delta}_\alpha] \right) &= -\frac{1}{\tau} \boldsymbol{a}^{(3,1)} + \boldsymbol{s}^{(3)}.
\end{aligned} \tag{A4}$$

Apparently $\boldsymbol{a}^{(0,1)} = \boldsymbol{a}^{(1,1)} = 0$. There is no unknown variable in the first two equations. In the last two equations, the time derivatives of $\boldsymbol{a}^{(2,eq)}$ and $\boldsymbol{a}^{(3,eq)}$ are reduced by the spatial derivatives with the aid of the first two equations in Eq. (A4):

$$\frac{\partial \boldsymbol{a}^{(2,eq)}}{\partial t} = -\frac{\partial}{\partial x} \cdot (\rho \boldsymbol{u}^3) - \left[ \boldsymbol{u} \frac{\partial}{\partial x} \rho \theta \right] + \boldsymbol{\delta} \frac{\partial}{\partial x} \cdot (\rho \boldsymbol{u}) - \boldsymbol{\delta} \left( \frac{2}{D} \rho \theta \left( \frac{\partial}{\partial x} \cdot \boldsymbol{u} \right) + \frac{\partial}{\partial x} \cdot (\rho \theta \boldsymbol{u}) \right)$$
$$- \boldsymbol{\delta}([\boldsymbol{s}^{(1)} \boldsymbol{u}] - \boldsymbol{s}^{(2)}): \boldsymbol{\delta}, \tag{A5}$$

$$\frac{\partial \boldsymbol{a}^{(3,eq)}}{\partial t} = -\frac{\partial}{\partial x} \cdot (\rho \boldsymbol{u}^4) + \left[ \boldsymbol{\delta} \frac{\partial}{\partial x_\mu} (\rho \boldsymbol{u} u_\mu) \right] + \left[ \boldsymbol{\delta} \frac{\partial}{\partial x} (\rho \theta) \right]$$
$$- \left[ \boldsymbol{\delta} \theta \frac{\partial}{\partial x} (\rho \theta) \right] - \left[ \boldsymbol{u}^2 \frac{\partial}{\partial x} (\rho \theta) \right] - \left[ \boldsymbol{\delta} \frac{\partial}{\partial x_\epsilon} (\rho \theta u_\epsilon \boldsymbol{u}) \right] \tag{A6}$$
$$- \frac{2}{D} \rho \theta \left[ \boldsymbol{\delta} \boldsymbol{u} \frac{\partial}{\partial x_\epsilon} u_\epsilon \right] + [\boldsymbol{s}^{(1)} \boldsymbol{u}^2] + \bar{\theta}[\boldsymbol{\delta} \boldsymbol{s}^{(1)}] + [\boldsymbol{u} \boldsymbol{\delta}]([-\boldsymbol{s}^{(1)} \boldsymbol{u} + \boldsymbol{s}^{(2)}]: \boldsymbol{\delta}).$$

Substituting Eq. (A5) and Eq. (A6) into eqn. (A4) the $\boldsymbol{a}^{(2,1)}$ and $\boldsymbol{a}^{(3,1)}$, as well as $\boldsymbol{a}^{(2,neq)}$ and $\boldsymbol{a}^{(3,neq)}$ are obtained:

$$\boldsymbol{a}^{(2,neq)} = -\tau\rho\theta\left(\left[\frac{\partial}{\partial \boldsymbol{x}}\boldsymbol{u}\right] - \frac{2}{D}\boldsymbol{\delta}\left(\frac{\partial}{\partial \boldsymbol{x}}\cdot\boldsymbol{u}\right)\right) + \tau\boldsymbol{\delta}([\boldsymbol{s}^{(1)}\boldsymbol{u}] - \boldsymbol{s}^{(2)}):\boldsymbol{\delta},$$

$$\boldsymbol{a}^{(3,neq)} = -\tau\left[\boldsymbol{\delta}\frac{\partial}{\partial x_\mu}\cdot(\rho\boldsymbol{u}u_\mu)\right] - \tau\rho\theta\left[\frac{\partial}{\partial \boldsymbol{x}}\theta\boldsymbol{\delta}\right] - \tau\rho\theta\frac{\partial}{\partial \boldsymbol{x}}\cdot[\boldsymbol{u}^2\boldsymbol{\delta}] \quad (A7)$$

$$+\tau\left[\boldsymbol{\delta}\frac{\partial}{\partial x_\epsilon}\rho\theta u_\epsilon\boldsymbol{u}\right] + \tau\rho\theta[\boldsymbol{\delta}\boldsymbol{u}]\frac{2}{D}\frac{\partial}{\partial x_\epsilon}u_\epsilon - \tau\boldsymbol{\Delta}',$$

Where

$$\boldsymbol{\Delta}' = -\boldsymbol{s}^{(3)} + [\boldsymbol{s}^{(1)}\boldsymbol{u}^2] + \bar{\theta}[\boldsymbol{\delta}\boldsymbol{s}^{(1)}] + [\boldsymbol{u}\boldsymbol{\delta}]([-\boldsymbol{s}^{(1)}\boldsymbol{u} + \boldsymbol{s}^{(2)}]:\boldsymbol{\delta}) \quad (A8)$$